\documentclass[cits]{PoS}

\title{Finite volume effects in SU(2) with two adjoint fermions}

\ShortTitle{Finite volume effects in SU(2) with two adjoint fermions}

\author{\speaker{Agostino Patella}\\
        CERN, Physics Department, 1211 Geneva 23, Switzerland\\
        E-mail: \email{agostino.patella@cern.ch}}

\author{Luigi Del Debbio%
        \thanks{The work of LDD, BL and AR is carried as part of the UKQCD collaboration and the DiRAC Facility jointly funded by STFC, the Large Facilities Capital Fund of BIS and Swansea University.}\\
        SUPA, School of Physics and Astronomy, University of Edinburgh, Scotland UK\\
        E-mail: \email{luigi.del.debbio@ed.ac.uk}}

\author{Biagio Lucini$^\dag$%
        \thanks{BL is supported by the Royal Society through the University Research Fellowship scheme and by STFC under contract ST/G000506/1.}\\
        College of Science, Swansea University, Singleton Park, Swansea SA2 8PP, UK\\
        E-mail: \email{B.Lucini@swansea.ac.uk}}

\author{Claudio Pica\\
        CP${}^3$-Origins \& Danish Institute for Advanced Study DIAS, University of Southern Denmark, Campusvej 55, DK-5230 Odense M, Denmark\\
        E-mail: \email{pica@cp3.sdu.dk}}

\author{Antonio Rago$^\dag$\\
        School of Computing and Mathematics, University of Plymouth, Plymouth PL4 8AA, UK\\
        E-mail: \email{antonio.rago@plymouth.ac.uk}}

\abstract{Many evidences from lattice simulations support the idea that SU(2) with two Dirac flavors in the adjoint representation (also called Minimal Walking Technicolor) is IR conformal. A possible way to see this is through the behavior of the spectrum of the mass-deformed theory. When fermions are massive, a mass-gap is generated and the theory is confined. IR-conformality is recovered in the chiral limit: masses of particles vanish in the chiral limit, while their ratios stay finite. In order to trust this analysis one has to relay on the infinite volume extrapolation. We will discuss the finite volume effects on the mesonic spectrum, investigated by varying the size of the lattice and by changing the boundary conditions for the fields.

\bigskip

\hspace*{9cm} CERN-PH-TH-2011-262\\
\hspace*{9cm} CP3-Origins-2011-035\\
\hspace*{9cm} DIAS-2011-27
}

\FullConference{XXIX International Symposium on Lattice Field Theory \\
		 July 10 - 16 2011\\
		 Squaw Valley, Lake Tahoe, California}

\begin{document}

\section{Introduction}

Many evidences collected with lattice simulations indicate that the SU(2) gauge theory with two Dirac fermions in the adjoint representation of the gauge group is inside the conformal window. A variety of investigation techniques has been used: the study of the spectrum in the mass-deformed theory in large volume~\cite{Catterall:2007yx, Catterall:2008qk, DelDebbio:2008tv, DelDebbio:2009fd, DelDebbio:2010hx, DelDebbio:2010hu, Bursa:2011ru, Hietanen:2008mr} and finite volume~\cite{DelDebbio:2010hu, Appelquist:2011dp}, the computation of the running coupling in the Schr\"odinger-functional scheme~\cite{Hietanen:2009az, Hietanen:2009zz, Karavirta:2011mv, Bursa:2009tj, Bursa:2009we, DeGrand:2011qd}, and finally Montecarlo RG methods \cite{Catterall:2011zf, Catterall:2011ce}.

We consider here the mass-deformed theory. The fermion mass generates a mass gap regardless of the theory being inside or outside of the conformal window. We observed in~\cite{DelDebbio:2009fd} that both the ratio $M_V/M_{PS}$ of the vector (V) and pseudoscalar (PS) isovector meson masses and the ratio $M_{PS}/\sqrt{\sigma}$ of the PS isovector meson mass and the square root of the string tension stay finite and different from zero, while each mass seems to go to zero. Moreover the lightest particle in the spectrum is the scalar glueball. This behavior is at odds with the predictions of spontaneous chiral symmetry breaking, and it is consistent with IR-conformality.

In case of IR-conformality, one expects~\cite{DelDebbio:2010hx, DelDebbio:2010ze, DelDebbio:2010jy} that all masses go to zero with the same power of the fermion mass, therefore ratios of masses would stay finite in the chiral limit. However it is crucial to notice that the power-law behavior is valid only in infinite volume.


Our goal is to study quantitatively the finite-size effects in the mass-deformed theory, in order to understand which of our previous results are safe.

\section{Two kinds of finite-size effects}

\begin{figure}[ht]
\centering
\includegraphics[width=0.67\textwidth]{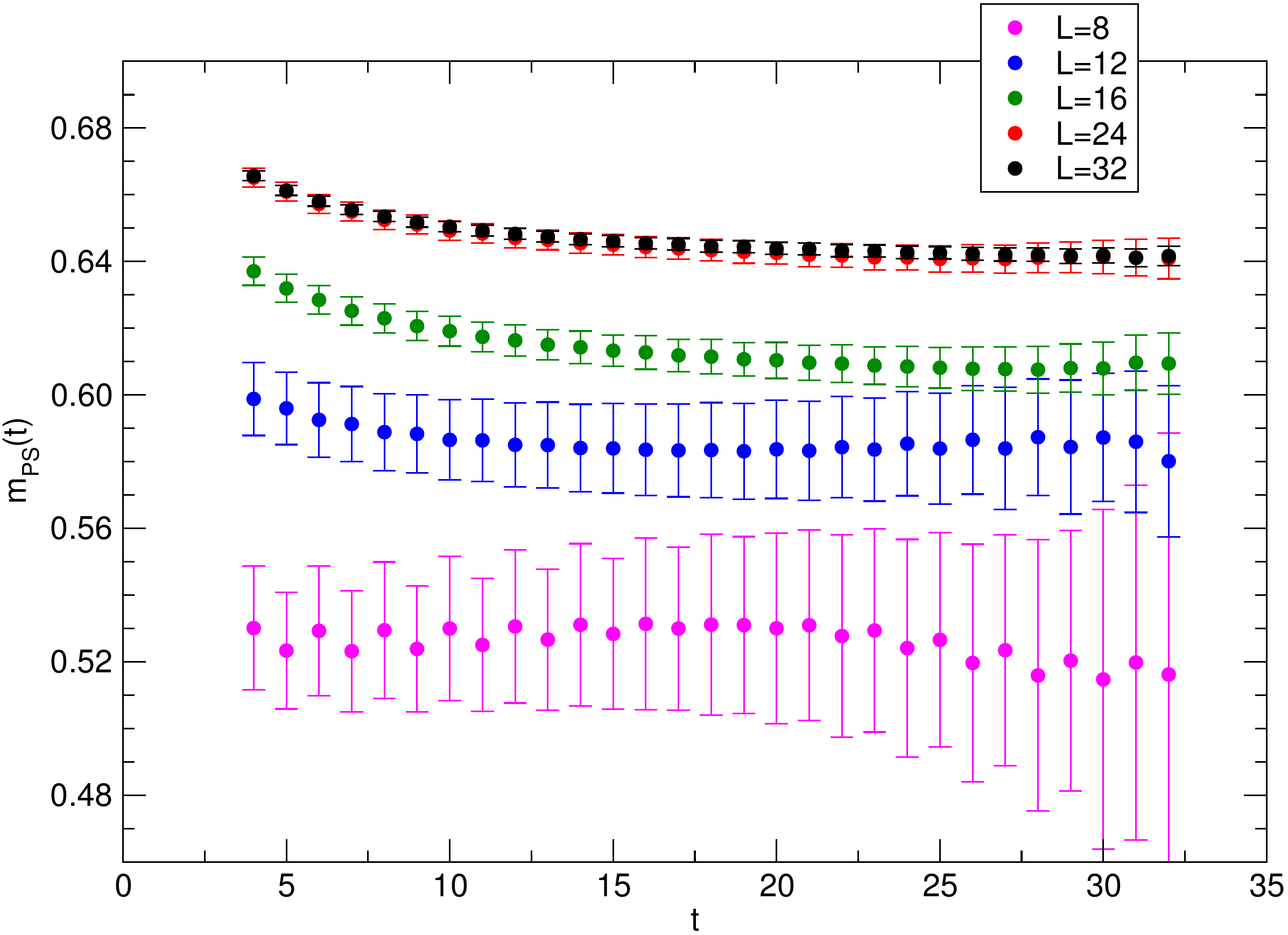}
\caption{Effective PS masses for different spatial boxes ($\beta=2.25$, $am_0=-1.15$, lattices $64 \times L^3$). A plateau in the effective masses is always visible. It is already clear that there is no change whatsoever while the lattice size is increased from $L=24$ to $L=32$.\label{fig:plateaux}}
\end{figure}

Measuring masses, one has to deal with two different finite-size effects because of the asymmetric treatment of the spatial and thermal directions. A finite spatial volume generates a genuine deformation of the spectrum. On the other hand, a finite thermal direction heats up the system: if no phase transition is crossed, the spectrum is not deformed but one can have difficulties in separating the ground state in the desired channel from the excited ones. In practice if the thermal direction is not large enough, one cannot see plateaux in the effective masses. There are two possible ways to get a better overlap with the lightest state: to increase the size of the thermal direction, or to use improved operators. The latter approach was exploited in~\cite{Bursa:2011ru}. However in this work we use point-like operators and we just choose a long enough temporal direction. The quality of the plateaux for the PS mass at $am_0 = -1.15$ (shown in fig.~\ref{fig:plateaux}), and the agreement with the results in~\cite{Bursa:2011ru} (whenever a comparison is possible) are clear signs that the finite-temperature effects are below our statistical errors. For the considered masses a thermal extent of $64$ lattice spacings is always enough to expose unambiguous plateaux.

\section{Mesons in a finite spatial box}

We study the effects of the finite spatial volume in two points in the parameter space. We always choose $\beta=2.25$ (which is the same value used in our previous works~\cite{DelDebbio:2008tv, DelDebbio:2009fd, DelDebbio:2010hx, DelDebbio:2010hu, Bursa:2011ru}). We focus on the two values for the bare mass $am_0=-1.15$ and $am_0=-1.05$ (corresponding respectively to PS masses of about $aM_{PS}\sim 0.64$ and  $aM_{PS}\sim 1.19$). We simulated on $64 \times L^3$ lattices with $L=8,12,16,24,32$. We will discuss here only the results corresponding to the lighter PS meson. However the same conlusions apply also to the heavier point.

In order to quantify finite-volume effects, we used two different techniques: \textit{(a)} simulations with usual periodic boundary conditions in space for the gauge and fermion fields on lattices with larger and larger spatial size, \textit{(b)} simulations with twisted boundary conditions in color space for the gauge and fermion fields \cite{'tHooft:1979uj}.

\begin{figure}[ht]
\centering
\includegraphics[width=0.48\textwidth]{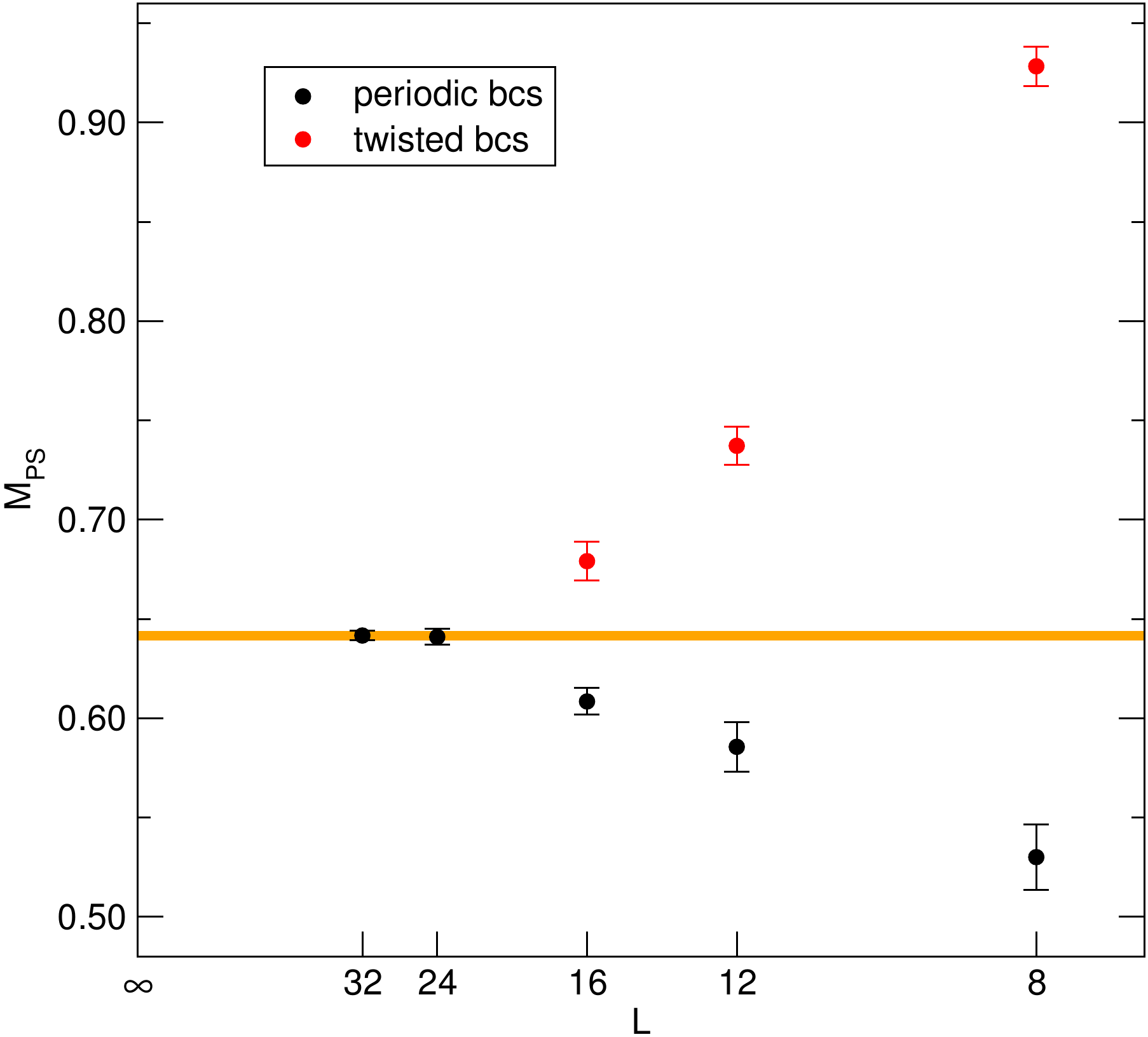}
\hspace{3mm}
\includegraphics[width=0.48\textwidth]{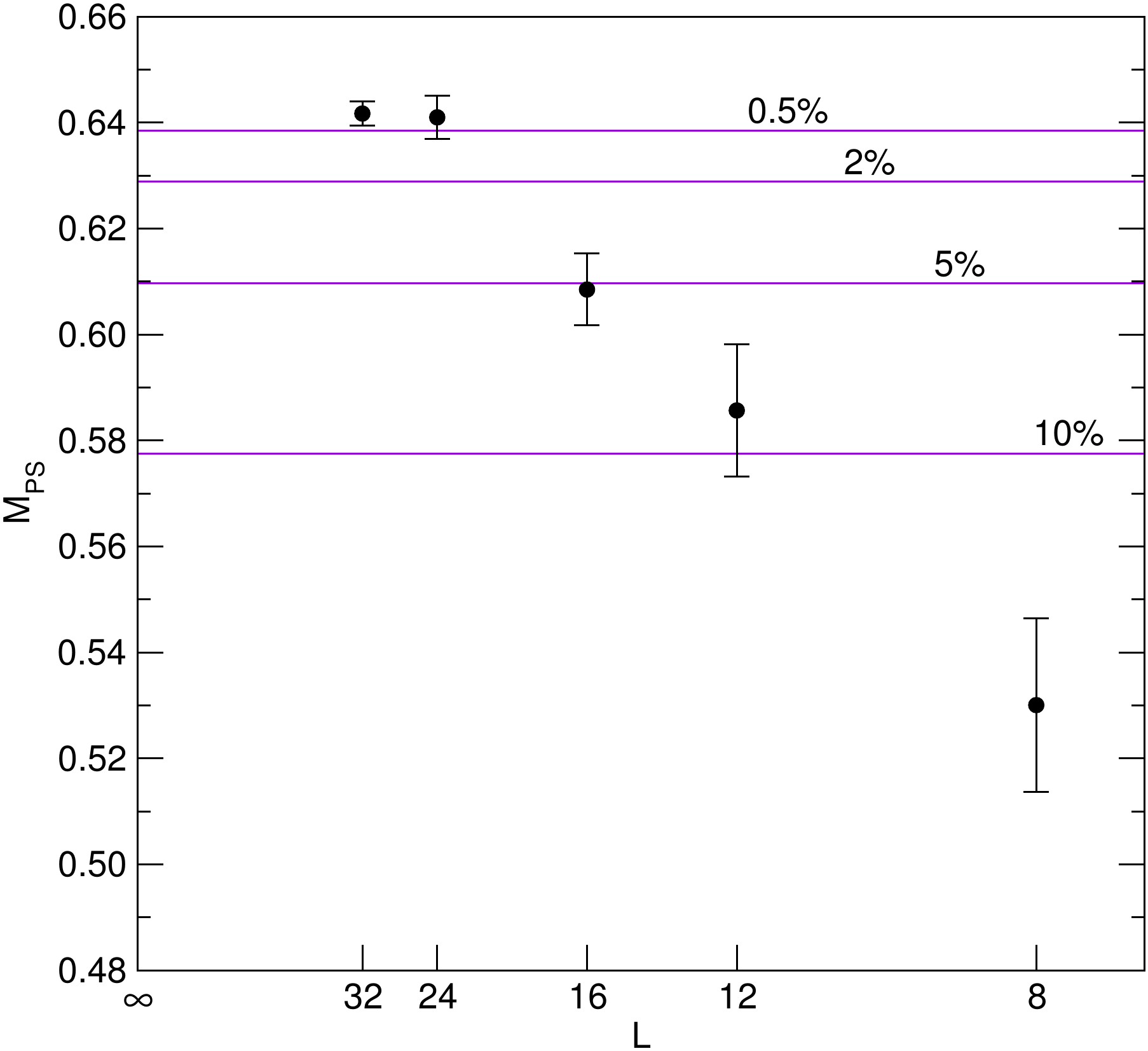}
\caption{($\beta=2.25$, $am_0=-1.15$, lattices $64 \times L^3$). \textit{Left pane.} PS mass with periodic (black) and twisted (red) boundary conditions. Changing boundary conditions changes the sign of the correction. The orange band is the result of a constant fit of the two points on the largest volume. \textit{Right pane.} PS mass with periodic boundary conditions. The horizontal lines correspond to a relative difference of $0.5\%$, $2\%$, $5\%$ and $10\%$ from the infinite-value estimate.\label{fig:mps}}
\end{figure}

In the particular case of the PS mass, changing the boundary conditions from periodic to twisted changes the sign of the finite-volume corrections (compare the black and red points in the left pane of fig.~\ref{fig:mps}). Hence this is a quite powerful method in order to estimate the order of magnitude of the finite-volume effects. However our results with twisted boundary conditions have to be seen as preliminary, since we do not reach the large-volume regime yet.

With periodic boundary conditions, there is no appreciable difference between the PS masses on the lattices with $L=24$ and $32$. Our statistical errors are here of the order of the $0.5\%$. We use the value of the PS mass $aM_{PS}=0.642(3)$ at $L=32$ as our estimate at infinite volume. We are in the position now to quantify the systematic error that we would make if we were choosing a smaller lattice. With $L=24$ we would have a systematic error of the same order of the statistical one. Reducing the spatial box to $L=16$ would give us a systematic error of about $5\%$, while $L=12$ gives a systematic error of about $10\%$. Assuming that the only relevant variable that determines the error due to a finite spatial volume is $M_{PS}L$ (as certainly it is in the case of IR-conformality), one can get the following estimates:
\begin{center}
\begin{tabular}{c|c}
$\quad M_{PS}L \quad$ & \textit{error due to the finite spatial box} \\
\hline
$8$ & $10\%$\\
$10$ & $5\%$\\
$13$ & $2\%$\\
$15$ & $0.5\%$
\end{tabular}
\end{center}

We learn immediately a very important lesson: \textit{if we were using the rule of thumb $M_{PS}L \sim 5$ that is commonly quoted for QCD, we would obtain finite-size systematics of about $20\%$ for the PS mass}. Even though we are simulating at fixed nonzero fermion mass, and therefore our system is confined (IR-conformality is recovered only in the chiral limit), finite-volume effects are much larger that the ones in QCD.

\begin{figure}[ht]
\centering
\includegraphics[width=0.75\textwidth]{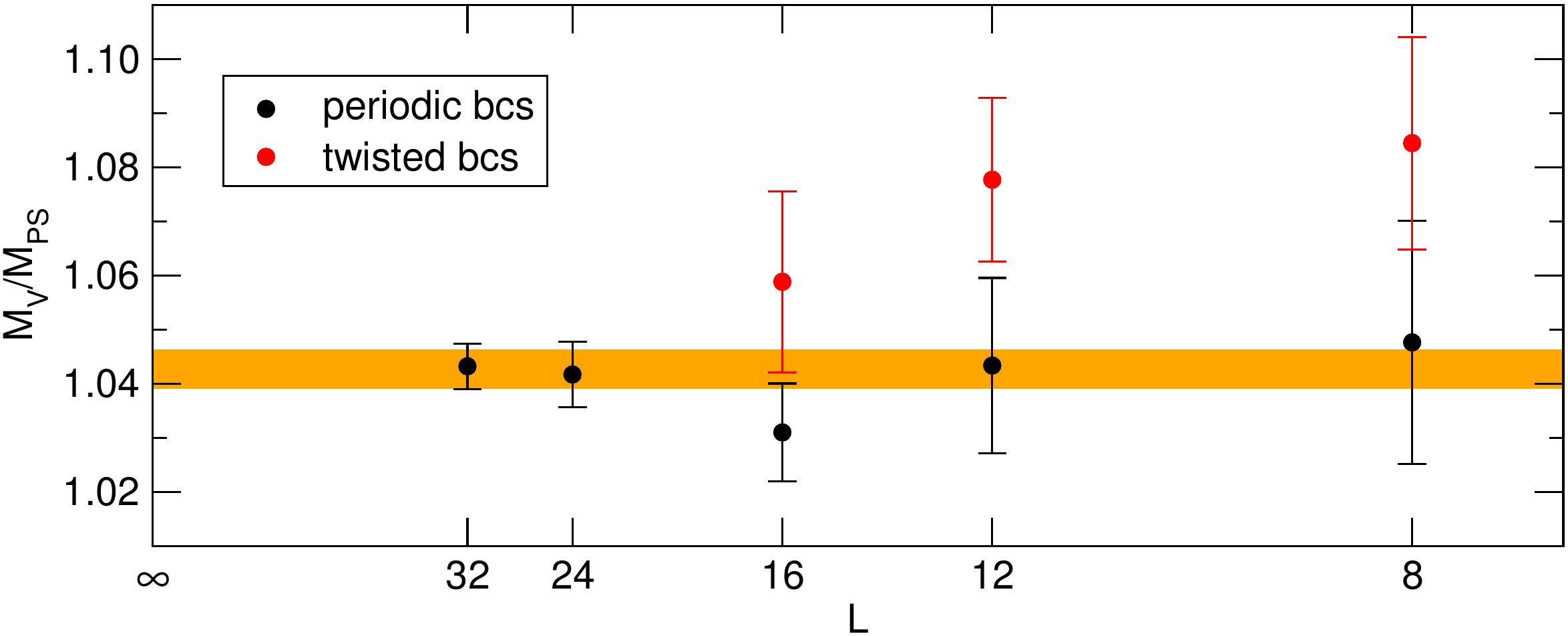}
\caption{($\beta=2.25$, $am_0=-1.15$, lattices $64 \times L^3$). Ratio of the V and PS masses with periodic (black) and twisted (red) boundary conditions. The determination with periodic boundary conditions has accidentally small finite-size corrections. The orange band is the result of a constant fit of all the points with periodic boundary conditions.\label{fig:ratio}}
\end{figure}

We also considered other mesonic observables (the PS decay constant, the vector and axial masses, several mass ratios), and $L=24$ corresponds always to the onset of the large-volume regime. It is of particular interest to notice that the finite-volume corrections seem to cancel out largely in the ratio $M_V/M_{PS}$ (with periodic boundary conditions, see fig.~\ref{fig:ratio}).

\section{Ratio of V and PS masses}

The exitence of a plateau in the $M_{V}/M_{PS}$ ratio as a function of the fermion mass was proposed as one of the crucial (but not the only!) signal for IR-conformality in~\cite{DelDebbio:2009fd}. We stress again that a plateau is expected on the basis of very basic RG analysis, \textit{provided that the volume can be considered almost infinite}. Which points in the plateau can be trusted to be at large enough volume?

\begin{figure}[ht]
\centering
\includegraphics[width=0.85\textwidth]{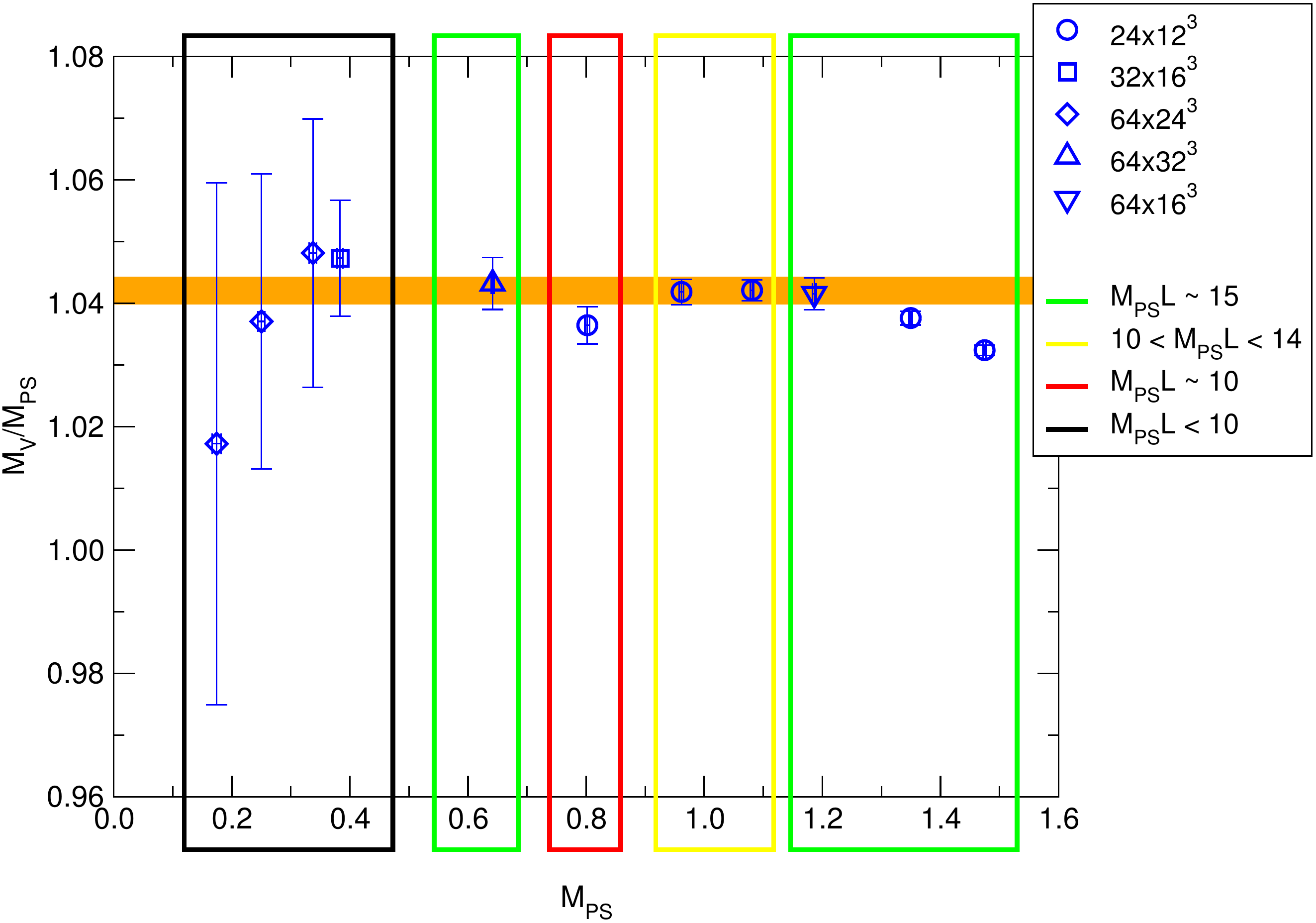}
\caption{Ratio of the V and PS masses at $\beta=2.25$. We choose the largest available volume for each fermion mass. The points enclosed in the green boxes are the ones that we can trust to be at almost infinite volume. A plateau can be trusted to exist in the range $0.64 \le aM_{PS} \le 1.19$.\label{fig:summary}}
\end{figure}

In fig.~\ref{fig:summary} we plot the ratio $M_{V}/M_{PS}$ at the largest available volume for each fermion mass. The points enclosed in the green boxes are the ones that we can trust to be at almost infinite volume. The points for which the product $M_{PS}L$ is in between $10$ and $14$ are enclosed in the yellow box. For the latter points the systematic error on the PS and V masses separately is in between $2\%$ and $5\%$, and are therefore fairly under control. However the relative error on the ratio itself is expected to be smaller. It is interesting to notice that the values of $M_{V}/M_{PS}$ at the two PS masses $aM_{PS}\sim 0.64$ and  $aM_{PS}\sim 1.19$ (both in the green region) are compatible. If one assumes the monotonicity of the $M_{V}/M_{PS}$ ratio, a plateau must exist in the range $0.64 \le aM_{PS} \le 1.19$. Unfortunately these masses are fairly heavy in terms of the cutoff.

Lighter points cannot be trusted a priori. It would be very useful to check the persistence of the plateau at a PS mass of about $aM_{PS} \sim 0.3$ against corrections due to the finite spatial box. We have already observed that, for the analyzed masses, the finite-size effects on the $M_{V}/M_{PS}$ ratio are much smaller that the ones on the masses separately. If this behavior persists at lighter masses, the infinite-volume value of the ratio might be reached at not too large volume. However how large the lattice should be cannot be said a priori, and a detailed study of finite size effects must be replicated at small masses. Control of the finite-size effects over the two masses separately would automatically imply control over the $M_{V}/M_{PS}$ ratio. For a PS mass of about $aM_{PS} \sim 0.3$, a $128 \times 48^3$ lattice would be surely enough to get systematics below $1\%$.

\section{Conclusions and comments}

The system we consider is the SU(2) gauge theory with two Wilson-Dirac fermions in the adjoint representation of the gauge group. Many evidences from lattice simulations indicate that this theory is IR-conformal in the chiral limit. We are interested to the corrections to the mesonic spectrum due to a finite spatial volume, at fixed nonzero fermion mass. A detailed study shows that $M_{PS}L$ has to be larger then $10$ to keep this correction below the $5\%$ for the PS mass, and larger than $15$ in order to keep it of the same order of our statistical errors (about $0.5\%$). These volumes are huge with respect to the ones needed for QCD (where the rule of thumb $M_{PS}L \sim 5$ is often adopted). Why does the $M_{PS}L$ product need to be so large? We can offer a simple interpretation based on our understanding of this particular theory.

In a theory with a mass gap, the finite-volume corrections to masses of stable particles are exponentially small $\sim e^{-mL}$. The mass $m$ controlling the exponential is related to the lightest state in the spectrum (for a quantitative analysis see~\cite{Koma:2004wz}, that is a generalization of the L\"uscher formula~\cite{Luscher:1985dn} for finite size effects to the case of two stable particles). Close to the chiral limit of a theory with spontaneous chiral symmetry breaking, the leading exponential is controlled by the pion mass, which is the lightest particle. However in the considered theory the lightest particle is the scalar glueball (let $M_G$ be its mass). Therefore the relevant product that one should consider is $M_G L$. How large $M_G L$ has to be in order to reach the large-volume regime? We can borrow some intuition from pure Yang-Mills. If we were compactifying only one direction, SU(2) pure Yang-Mills would undergo a deconfinement transition at about $M_G L \sim 5$. Since we compactify all three spatial directions, no phase transition occurs. Still, the effective potential for the spatial Polyakov loop changes its shape (from one to two minima) generating a crossover which is a pure finite-volume effect. In order to reach the large-volume regime one needs to be at $M_G L \gtrsim 5$. In the considered theory the PS meson is twice as heavier as the lightest glueball, which gives $M_{PS} L \gtrsim 10$ for the large-volume regime. Although this is a rough estimate, it captures the correct order of magnitude.

We also want to comment on the sign of the correction to the PS mass due to a finite spatial box. Squeezing the box, the PS meson becomes lighter. One could argue that when the box is squeezed the mass should increase because of the indetermination principle. However this argument would apply only to wave functions with Dirichlet conditions at the boundary of the box. For fields with periodic boundary conditions the analysis is more complicated. Both the L\"uscher formula~\cite{Luscher:1985dn} and its generalization in~\cite{Koma:2004wz} show that the leading correction due to a finite box is negative. Therefore the PS meson has to become lighter as the box is squeezed, unless some special mechanism (like spontaneous chiral symmetry breaking) kills the leading negative term.

\end{document}